\documentstyle[12pt]{article}

\begin{document}

\hfill{EFUAZ FT-98-57}

\vspace*{1cm}

\begin{center}
{\bf Question. \,\, Negative Group Delays
and Action-at-a-Distance}\\
(submitted to the section ``Questions and Answers"
of ``Am. J. of Phys.")\\
\end{center}

\bigskip
\bigskip

In the recent issue of {\it American Journal of Physics}
M. W. Mitchell and R. Y. Chiao$^1$
discuss the causality and negative group delays
in the series of the experiments on the superluminal
propagation of electromagnetic waves (see references
therein).

We have  a question. On p. 14 the authors wrote
``it is not the group velocity, but rather
the front velocity that must be no greater than $c$
by Einstein causality"; on the other hand on p. 17
the authors wrote ``...the front...
reaches the input and the output of each
amplifier at the {\bf same} time". We do not understand
how one should reconcile these two expressions.

Even if one assumes that the time of flight
for light to cross from the input  to the
output  is completely negligible on the scale
of the typical time scales for the pulses and
``the front reaches the input
and the output of each amplifier at the [{\bf almost}]
same time", it is also not  comprehensible for us
(in the framework of the Einstein claim
about the $c$ as the limiting velocity)
the following expression [p.18, Eq. (12)]:
``...the output depends only
on the {\bf present} and past values of the input
(and not on future values)".
Theoretically we agree (we are theorists).
But, in our opinion, this statement
(in comparison with the previous statements)
also can cause misunderstandings: if
the output depends on the {\bf  present}
values of the input (see the term $V_{in} (t)$
in Eq. (12), $t$ is the same time, not retarded!),
does the above statement signify that
{\bf there is} information in the output
which comes with the ``{\bf infinite} velocity",
thus approving the action at a distance,$^2$
does not it?

In this context we wonder if the formalism of the
causal Green function
($V_{out} (t) = V_{in} (t)
+\mbox{retarded terms}$, Eq. (12)$^1$) contradicts
with what Einstein and his successors told us?
We wonder, if it is possible to construct some
device in order to check experimentally,
whether the front comes at the {\bf same time}
or at the {\bf almost same} time.

\bigskip
\bigskip

\noindent
$^1$ M. W. Mitchell and R. Y. Chiao,
``Causality and negative group delay in a simple
bandpass amplifier", Am. J. Phys. {\bf 66} (1),
14-19 (1998)\\

\noindent
$^2$ A. E. Chubykalo and R. Smirnov-Rueda,
``Convection dispacement current and generalized form
of Maxwell-Lorentz equations", Mod. Phys. Lett. A{\bf 12} (1),
1-24 (1997)\\

\bigskip
\bigskip

\noindent
Andrew E. Chubykalo, Valeri V. Dvoeglazov\\
{\it Escuela de F\'{\i}sica\\
Universidad Aut\'onoma de Zacatecas\\
Apartado Postal C-580\\
Zacatecas 98068 Zac. M\'exico}\\

\end{document}